\documentclass[prb,twocolumn]{revtex4}
\usepackage{amssymb}
\usepackage{amsmath}
\usepackage{amsfonts}
\usepackage{graphicx}
\usepackage{float}
\begin{document}
\author{A. E. Koshelev}
\affiliation{Materials Science Division, Argonne National Laboratory, Argonne, Illinois 60439}
\title{Electrodynamics of the Josephson vortex lattice in high-temperature superconductors}

\begin{abstract}
We studied the response of the Josephson vortex lattice in layered
superconductors to the high-frequency c-axis electric field. We found a simple
relation connecting the dynamic dielectric constant with the perturbation of
the superconducting phase, induced by oscillating electric field. Numerically
solving equations for the oscillating phases, we computed the frequency
dependences of the loss function at different magnetic fields, including
regions of both dilute and dense Josephson vortex lattices. The overall
behavior is mainly determined by the c-axis and in-plane dissipation
parameters, which are inversely proportional to the anisotropy. The cases of
weak and strong dissipations are realized in
$\mathrm{Bi_{2}Sr_{2}CaCu_{2}O_{x}}$ and underdoped $\mathrm{YBa_{2}Cu_{3}
O_{x}}$, respectively. The main feature of the response is the
Josephson-plasma-resonance peak. In the weak-dissipation case additional
satellites appear in the dilute regime in the higher-frequency region due to
the excitation of the plasma modes with the wave vectors set by the lattice
structure. In the dense-lattice limit the plasma peak moves to a higher
frequency and its intensity rapidly decreases, in agreement with experiment and
analytical theory. The behavior of the loss function at low frequencies is well
described by the phenomenological theory of vortex oscillations. In the case of
very strong in-plane dissipation an additional peak in the loss function
appears below the plasma frequency. Such peak has been observed experimentally
in underdoped $\mathrm{YBa_{2}Cu_{3} O_{x}}$. It is caused by the frequency
dependence of the in-plane contribution to losses rather than a definite mode
of phase oscillations.

\end{abstract}
\date{\today }
\maketitle

\section{Introduction}

The Josephson plasma resonance (JPR)\cite{art93,TKTPRB94,bmtPRL95} is one of
the most prominent manifestations of the intrinsic Josephson effect in layered
superconductors.\cite{IJJReviews,YurgensReview} It has been established as a
valuable tool to study the intrinsic properties of
superconductors\cite{GaifullinPRL99,DordevicPRB05} and it was extensively used
to probe different states of vortex
matter.\cite{OngJPR,MatsudaJPR,KadowJPR,ShibauchiPRL99,vdBeek,DulicPRL01,Thorsmolle}
The value of the JPR frequency depends on the anisotropy factor and it ranges
widely for different compounds, from several hundred gigahertz in
$\mathrm{Bi_{2}Sr_{2}CaCu_{2}O_{x}}$ (BSCCO) to several terahertz in
underdoped $\mathrm{YBa_{2}Cu_{3} O_{x}}$ (YBCO).
%These two material will be mostly discussed in this manuscript.

An interesting issue is the influence of the magnetic field applied along the
layer direction on the high-frequency response and, in particular, on the JPR.
Such field forms the lattice of Josephson vortices (JVL). The anisotropy
factor, $\gamma$, and the interlayer periodicity, $s$, set the important field
scale, $B_{\mathrm{cr}}=\Phi_{0}/(2\pi\gamma s^{2})$. At $B<B_{\mathrm{cr}}$
the Josephson vortices are well separated and form a dilute lattice. When the
magnetic field exceeds $B_{\mathrm{cr}}$ the Josephson vortices homogeneously
fill all layers (dense-lattice
regime).\cite{BulClemPRB91,KoshGrState,NonomHuPRB06} The crossover field ranges
from $\sim$0.5 tesla for BSCCO to $\sim$10 tesla for underdoped YBCO. The JVL
state is characterized by a very rich spectrum of dynamic properties. In
particular, the transport properties of the JVL in BSCCO have been extensively
studied by several experimental groups.\cite{Lee,Hechtfischer,Latyshev}

The high-frequency response in the magnetic fields along the layers has been
studied experimentally in BSCCO using the microwave absorption in cavity
resonators \cite{KakeyaPRB05}, in $\mathrm{La_{2-x}Sr_{x}CuO_{4}}$
\cite{DordevicPRB05}, and in underdoped YBCO \cite{KojimaJLTP03,LaForge} by the
infrared reflection spectroscopy. A detailed comparison between the behaviors
of the high-frequency response in the in-plane magnetic field for these
compounds has been made recently by LaForge \emph{et al.}\cite{LaForge} In
BSCCO, two resonance absorption peaks have been observed.\cite{KakeyaPRB05} The
upper-resonance frequency increases with the field and approaches the JPR
frequency at small fields, while the lower-resonance frequency decreases with
the field and approaches approximately half of the JPR frequency at small
fields. In the underdoped YBCO\cite{KojimaJLTP03,LaForge}, the JPR peak in the
loss function rapidly loses its intensity with increasing the field, while the
resonance frequency either does not move or slowly increases with the field. In
addition, another wide peak emerges at a smaller frequency and its intensity
increases with increasing field.

Several theoretical approaches have been used to describe the
response of the JVL state to the oscillating electric field. A
phenomenological vortex-oscillation theory has been proposed by
Tachiki \textit{et al.} \cite{TKTPRB94}. This theory describes the
response of the JVL at small frequencies and fields, in terms of
phenomenological vortex parameters, mass, viscosity coefficient, and
pinning spring constant. This approach is expected to work at
frequencies much smaller then the JPR frequency, i.e., it can not be
used to describe the JPR peak itself.

On the other hand, the plasmon spectrum at high magnetic fields, in the
dense-lattice limit, has been calculated by Bulaevskii \textit{et al.}
\cite{BulPRB97} It was found that the plasma mode at zero wave vector increases
proportionally to the magnetic field, $\omega_{p}(B)=\omega
_{p}(0)B/(2B_{\mathrm{cr}})$. This prediction describes very well the behavior
of the high-frequency mode in BSCCO.\cite{KakeyaPRB05} A more quantitative
numerical study of the JVL plasma modes in the dense-lattice regime has been
done by Koyama.\cite{KoyamaPRB03} He reproduced the mode, linearly growing with
field, and also found additional modes at smaller frequencies.

In this paper we develop a quantitative description of the high-frequency
response for a homogenous layered superconductor valid for whole range of
frequencies and fields. We relate the dynamic dielectric constant via simple
averages of the oscillating phases. The high-frequency response is mainly
determined by the c-axis and in-plane dissipation parameters, which are
inversely proportional to the anisotropy. Analytical results for the dynamic
dielectric constant and loss function can be derived in limiting cases, at
small fields and frequencies (vortex-oscillation regime) and at high fields, in
the dense-lattice regime. Numerically solving dynamic equations for the
oscillating phases, we studied applicability limits of the approximate
analytical results and investigated the influence of the dissipation parameters
on the shape of the loss function. Computing the oscillating phases for
different vortex lattices, we traced the field evolution of the loss function
with increasing magnetic field for the cases of weak and strong dissipations,
typical for BSCCO and underdoped YBCO, respectively.

\section{Dynamic phase equations, dielectric constant, and loss function}

The equations describing phase dynamics in layered superconductors can be
derived from Maxwell's equations expressing fields and currents in terms of the
gauge invariant phase difference between the layers $\theta_{n}=\phi
_{n+1}-\phi_{n}-\left(  2\pi s/\Phi_{0}\right)  A_{z}$. These equations have
been presented in several different forms.\cite{Inductive} We will use them in
the form of coupled equations for the phase differences and magnetic field when
charging effects can be neglected, see, e.g., Refs.\
\onlinecite{ArtJETPL97,KoshAranPRB01},
\begin{subequations}
\begin{align}
& \frac{\sigma_{c}\Phi_{0}}{2\pi csj_{J}}\frac{\partial\theta_{n}}{\partial
t}\!+\!\sin\theta_{n}\!+\!\frac{1}{\omega_{p}^{2}}\frac{\partial^{2}\theta_{n}
}{\partial t^{2}}\!-\!\frac{c}{4\pi j_{J}}\frac{\partial B_{n}}{\partial x}
\!=\!\frac{1}{4\pi j_{J}}\frac{\partial\mathcal{D}_{z}}{\partial
t},\label{PhaseEq1}\\
& \left(  \frac{4\pi\sigma_{ab}}{c^{2}}\frac{\partial}{\partial
t}+\frac {1}{\lambda_{ab}^{2}}\right)  \left(  \frac{\Phi_{0}}{2\pi
s}\frac {\partial\theta_{n}}{\partial x}-B_{n}\right)
=-\frac{\nabla_{n}^{2} B_{n} }{s^{2}},\label{PhaseEq2}
\end{align}
where the magnetic field is along the $y$ axis, $\sigma_{ab}$ and $\sigma_{c}$
are the components of the quasiparticle conductivity, $\lambda_{ab}$ and
$\lambda_{c}$ are the components of the London penetration depth, $j_{J}
=c\Phi_{0}/(8\pi^{2}s\lambda_{c}^{2})$ is the Josephson current density,
$\omega_{p}=c/\sqrt{\varepsilon_{c}}\lambda_{c}$ is the plasma frequency,
$\mathcal{D}_{z}$ is the external electric field, and
$\nabla_{n}^{2}B_{n}\equiv B_{n+1}+B_{n-1}-2B_{n}$. Neglecting charging effects
\cite{charging}, the local electric field is connected with the phase
difference by the Josephson relation,
\end{subequations}
\begin{equation}
E_{z}\approx\frac{\Phi_{0}}{2\pi cs}\frac{\partial\theta_{n}}{\partial
t}.\label{JosRel}
\end{equation}
The average magnetic induction inside the superconductor, $B_{y}$,
fixes the average phase gradient, $\left\langle
\partial\theta_{n}^{(0)}/\partial x\right\rangle =2\pi
sB_{y}/\Phi_{0}$.

To facilitate analysis, we use a standard transformation to the reduced
variables,
\[
x/\lambda_{J}\rightarrow x,\ \  t\omega_{p}\rightarrow t,\ \
h_{n}=2\pi\gamma s^{2}B_{n}/\Phi_{0},
\]
with $\lambda_{J}=\gamma s$, and introduce the dimensionless
parameters
\[
l=\frac{\lambda_{ab}}{s},\ \nu_{c}=\frac{4\pi\sigma_{c}}{\varepsilon_{c}
\omega_{p}},\ \nu_{ab}=\frac{4\pi\sigma_{ab}\lambda_{ab}^{2}\omega_{p}}{c^{2}
}.
\]
It is important to note that both damping parameters, $\nu_{c}$ and $\nu_{ab}
$, roughly scale inversely proportional to the anisotropy factor $\gamma$,
meaning that \emph{the effective damping is stronger in less anisotropic
materials}. In particular, the cases of weak and strong dissipation are
realized in BSCCO and underdoped YBCO, respectively. Due to the d-wave pairing
in the high-temperature superconductors, both dissipation parameters do not
vanish at $T\rightarrow0$. Another important feature of the high-temperature
superconductors is that the in-plane dissipation is typically much stronger
than the c-axis dissipation, $\nu_{ab}\gg\nu_{c}$.\cite{LatyshevKBPRB03} This
is a consequence of a rapid decrease of the in-plane scattering rate with
decreasing temperature, which manifests itself as a large peak in the
temperature dependence of the in-plane quasiparticle
conductivity.\cite{quasiYBCO,quasiBSCCO}

Assuming an oscillating external field and using a complex presentation,
$\mathcal{D}_{z}(t)=\mathcal{D}_{z}\exp(-i\omega t)$, we obtain for small
oscillations
\begin{subequations}
\begin{align}
\left(  -i\nu_{c}\tilde{\omega}+C_{n}(x)-\tilde{\omega}^{2}\right)  \theta
_{n}-l^{2}\frac{\partial h_{n}}{\partial x}  &  =\frac{i\omega}{4\pi j_{J}
}\mathcal{D}_{z}\label{OscPhEq1}\\
\frac{\partial\theta_{n}}{\partial x}-h_{n}+\frac{l^{2}}{1-i\nu_{ab}
\tilde{\omega}}\nabla_{n}^{2}h_{n}  &  =0\label{OscPhEq2}
\end{align}
where $\tilde{\omega}=\omega/\omega_{p}$, $C_{n}(x)\equiv\cos[\theta_{n}
^{(0)}(x)]$, and the static phases, $\theta_{n}^{(0)}(x)$, are determined by
the following reduced equations
\end{subequations}
\begin{equation}
\frac{\partial^{2}\theta_{n}^{(0)}}{\partial x^{2}}+\left(  -\frac{1}{l^{2}
}+\Delta_{n}\right)  \sin\theta_{n}^{(0)}=0\label{StaticPhaseDifEqRed}
\end{equation}
with the addition condition $\left\langle \partial\theta_{n}
^{(0)}/\partial x\right\rangle =h=2\pi\gamma s^{2}B_{y}/\Phi_{0}$.
We introduce the reduced oscillating phase $\vartheta_{n}$,
\[
\theta_{n}=\vartheta_{n}\frac{\omega_{p}\mathcal{D}_{z}}{4\pi j_{J}},
\]
for which we can derive from Eqs.\ (\ref{OscPhEq1}) and (\ref{OscPhEq2}) the
following reduced equation
\begin{align}
-  &  \frac{\partial^{2}\vartheta_{n}}{\partial x^{2}}+\left(  \frac{1}{l^{2}
}-\frac{1}{1-i\nu_{ab}\tilde{\omega}}\nabla_{n}^{2}\right) \nonumber\\
&  \times\left[  C_{n}(x)-\tilde{\omega}^{2}-i\nu_{c}\tilde{\omega}\right]
\vartheta_{n}=-\frac{i\tilde{\omega}}{l^{2}}.\label{ReducOscPh}
\end{align}
From the Josephson relation (\ref{JosRel}) we find $E_{z}=\left(
-i\tilde{\omega}/\varepsilon_{c}\right) \overline{\vartheta}\mathcal{D}_{z}$,
meaning that the dynamic dielectric constant is connected with the average
reduced oscillating phase, $\overline{\vartheta}$, by a simple relation
\begin{equation}
\varepsilon_{c}(\tilde{\omega})=-\varepsilon_{c}/(i\tilde{\omega}
\overline{\vartheta}).\label{DielCnstOscPhase}
\end{equation}

Consider the case of a zero magnetic field first. In this case $\theta_{n}
^{(0)}(x)=0$ and the oscillating phase is given by
\begin{equation}
\vartheta_{n}=\overline{\vartheta}=\frac{-i\tilde{\omega}}{1-\tilde{\omega
}^{2}-i\nu_{c}\tilde{\omega}}.\label{OscPhaseZeroB}
\end{equation}
In this case Eq.\ (\ref{DielCnstOscPhase}) gives well-known results for the
dynamic dielectric constant, $\varepsilon
_{c0}(\omega)\equiv\mathcal{D}_{z}/E_{z}$, and the loss function,
$L_{0}(\omega)=\operatorname{Im}[-1/\varepsilon_{c0}(\omega)]$ at the zero
magnetic field, which we present in real units,
\begin{align}
\varepsilon_{c0}(\omega)  &  =\varepsilon_{c}-\frac{\varepsilon_{c}\omega
_{p}^{2}}{\omega^{2}}+\frac{4\pi i\sigma_{c}}{\omega},\label{DynEpsB0}\\
L_{0}(\omega)  &
=\frac{4\pi\omega^{3}\sigma_{c}/\varepsilon_{c}^{2}}{\left(
\omega^{2}-\omega_{p}^{2}\right)  ^{2}+\left(  4\pi\omega\sigma_{c}
/\varepsilon_{c}\right)  ^{2}}.\label{LB0}
\end{align}
The zero-field loss function has a peak at the Josephson plasma
frequency with width determined only by the c-axis quasiparticle
conductivity.

Consider now the case of a finite magnetic field applied in the layer
direction. Such magnetic field generates the lattice of the Josephson vortices,
which is described by the static phase distribution, $\theta_{n}^{(0)}(x)$,
obeying Eq.~(\ref{StaticPhaseDifEqRed}). The oscillating phase, in turn, is
fully determined by this phase via the spatial distribution of the average
cosine, $C_{n}(x)$. Averaging Eq.~(\ref{StaticPhaseDifEqRed}), we obtain an
obvious identity $\langle \sin\theta_{n}^{(0)}\rangle =0$, indicating that
there is no average current in the ground state. In the field range $B_{y}\gg
\Phi_{0}/2\pi\lambda_{ab}\lambda_{c}$ the term $1/l^{2}$ can be neglected. In
this limit it is more convenient to operate with the in-plane phases $\phi_{n}
^{(0)}(x)$, defined by the relation $\theta_{n}^{(0)}=\phi_{n+1}^{(0)}-\phi
_{n}^{(0)}+hx$, which obey the following equation
\begin{align}
\frac{\partial^{2}\phi_{n}^{(0)}}{\partial x^{2}}  &  +\sin\left(  \phi
_{n+1}^{(0)}-\phi_{n}^{(0)}+hx\right) \nonumber\\
&  -\sin\left(  \phi_{n}^{(0)}-\phi_{n-1}^{(0)}+hx\right)
=0.\label{StatPhEq}
\end{align}
Due to the translational invariance, every solution $\phi_{n}^{(0)}(x)$
generates a continuous family of solutions $\phi_{n}^{(0)}(x-u)$ corresponding
to the lattice shifts $u$. In particular, the phase change for a small
displacement is given by $\delta\phi_{n}\!=\!-u\partial\phi_{n}^{(0)}/\partial
x$. Taking the $x$-derivative of Eq. (\ref{StatPhEq}), we find that
$\partial\phi _{n}^{(0)}/\partial x$ obeys the following equation
\begin{align}
&  \frac{\partial^{3}\phi_{n}^{(0)}}{\partial x^{3}}+C_{n}\frac{\partial
\phi_{n+1}^{(0)}}{\partial x}+C_{n-1}\frac{\partial\phi_{n-1}^{(0)}}{\partial
x}\nonumber\\
&  -\left(  C_{n}+C_{n-1}\right)  \frac{\partial\phi_{n}^{(0)}}{\partial
x}=-\nabla_{n}C_{n}h\label{StatPhDerEq}
\end{align}
with $C_{n}=\cos\left(  \phi_{n+1}^{(0)}-\phi_{n}^{(0)}+hx\right)$
and $\nabla_{n}C_{n}=C_{n}-C_{n-1}$. The condition $\left\langle
\partial \sin\theta_{n}^{(0)}/\partial x\right\rangle =0$ gives
another useful identity for $\partial\phi_{n}^{(0)}/\partial x$
\begin{equation}
\overline{C_{n}\left(  \frac{\partial\phi_{n+1}^{(0)}}{\partial x}
-\frac{\partial\phi_{n}^{(0)}}{\partial x}\right)  }=-Ch.\label{AverPhDer}
\end{equation}

We now proceed with the analysis of the dynamic phase equation
(\ref{ReducOscPh}). Averaging this equation, we obtain the following relation
\[
\left(  i\nu_{c}\omega+\omega^{2}\right)  \overline{\vartheta}-\overline
{C_{n}(x)\vartheta_{n}}=i\omega.
\]
Splitting $\vartheta_{n}(x)$ into the average and oscillating-in-space parts,
$\vartheta_{n}(x)=\overline{\vartheta}\left(  1+w_{n}(x)\right)  $, with
$\overline{w_{n} (x)}=0$, we obtain
\[
\overline{\vartheta}=\frac{i\tilde{\omega}}{\omega^{2}-C-\overline
{C_{n}(x)w_{n}}+i\nu_{c}\tilde{\omega}}.
\]
Using relation (\ref{DielCnstOscPhase}), we obtain the following
result for the dielectric constant
\begin{equation}
\frac{\varepsilon_{c}}{\varepsilon_{c}(\tilde{\omega})}=\frac{\tilde{\omega
}^{2}}{\tilde{\omega}^{2}-C-\overline{C_{n}(x)w_{n}}+i\nu_{c}\tilde{\omega}
}\label{DielConstJVL}.
\end{equation}
Therefore, the dynamic dielectric constant is fully determined by the two
simple averaged quantities, the static average cosine, $C$, and the average
including the spatial distribution of the oscillating phase,
$\overline{C_{n}(x)w_{n}}$.

Introducing again the in-plane oscillating phases,
$w_{n}=\phi_{n+1}-\phi_{n} $, and neglecting terms of the order of
$1/l^{2}$, we derive the following equation
\begin{align}
\frac{\partial^{2}\phi_{n}}{\partial x^{2}}  &  +\frac{1}{1-i\nu_{ab}
\tilde{\omega}}\nabla_{n}\left[  C_{n}(x)-\tilde{\omega}^{2}-i\nu_{c}
\tilde{\omega}\right]  \nabla_{n}\phi_{n}\nonumber\\
&  =-\frac{\nabla_{n}C_{n}(x)}{1-i\nu_{ab}\tilde{\omega}}.\label{DynPhEq}
\end{align}
The oscillating phase has the same symmetry properties as the vortex lattice.
Comparing this equation with Eq. (\ref{StatPhDerEq}), we see that in the limit
$\omega\rightarrow0$ the solution is given by
\begin{equation}
\phi_{n}\rightarrow\frac{1}{h}\frac{\partial\phi_{n}^{(0)}}{\partial
x}.\label{StatLimSol}
\end{equation}
This solution corresponds to the homogeneous lattice shift. As
follows from Eq.\ (\ref{AverPhDer}), the combination $C+\overline
{C_{n}(x)w_{n}}=C+\overline{C_{n}(x)\left(
\phi_{n+1}-\phi_{n}\right)  }$, which determines the dielectric
constant (\ref{DielConstJVL}), vanishes in this limit. This property
is a consequence of the translational invariance and it is only true
in the absence of pinning of the vortex lattice.

In summary, to find the dynamic dielectric constant at a given field, one has
to find first the static phase from Eq.\ (\ref{StatPhEq}) assuming a definite
vortex-lattice structure and compute the average cosine, $C=\left\langle
\cos\left( \phi_{n+1}^{(0)}-\phi_{n}^{(0)}+hx\right) \right\rangle $. After
that, one has to solve the dynamic equation (\ref{DynPhEq}) and compute the
average $\overline{C_{n}(x)w_{n}}\!=\!\overline{C_{n}(x)\left(
\phi_{n+1}\!-\!\phi _{n}\right)  }$. These two averages completely determine
$\varepsilon _{c}(\omega)$ via Eq.~(\ref{DielConstJVL}). In the following
sections, we will consider regimes in which analytical solutions are possible,
the high-field limit and the vortex-oscillation regime at small frequencies.
Then, we will present the results of the numerical analysis in the full field
and frequency range in the cases of weak dissipation (BSCCO) and large
dissipation (underdoped YBCO).

\section{High-field regime \label{sec:HighField}}

In this section we consider the high-field regime, $h\gg1$, corresponding to
the dense-lattice limit in which all interlayer junctions are homogeneously
filled with vortices. This regime allows for the full analytical description
using an expansion with respect to the Josephson currents. In particular, the
spectrum of the plasma modes and their damping in this limit have been found by
Bulaevskii \textit{et al.}\cite{BulPRB97} We use the same approach to derive
the dynamic dielectric constant.

The static phase solution at high fields, describing the triangular lattice, is
given by
\begin{equation}
\phi_{n}^{(0)}\approx\frac{\pi n(n-1)}{2}+\frac{2}{h^{2}}\sin\left(  hu+\pi
n\right)  .\label{StatPhHighH}
\end{equation}
Using this distribution, we compute the average cosine
\begin{equation}
C\approx\overline{\cos\left[  hu+\pi n-\frac{4}{h^{2}}\sin\left(
hu+\pi n\right)  \right] }\approx\frac{2}{h^{2}}.\label{AverCosHigh}
\end{equation}
At high fields we can neglect rapidly oscillating $C_{n}(x)$ in the right hand
side of Eq. (\ref{DynPhEq}). This allows us to obtain the solution
\begin{equation}
\phi_{n}=-\frac{C_{n}(x)/2}{\tilde{\omega}^{2}-\left(  1-i\nu_{ab}
\tilde{\omega}\right)  h^{2}/4+i\nu_{c}\tilde{\omega}},\label{DynPhaseHigh}
\end{equation}
compute the average
\[
\overline{C_{n}(x)\left(  \phi_{n+1}-\phi_{n}\right)  }=\frac{1/2}
{\tilde{\omega}^{2}-\left(  1-i\nu_{ab}\tilde{\omega}\right)
h^{2}/4+i\nu _{c}\tilde{\omega}},
\]
and, finally, obtain the high-field limit of $\varepsilon_{c}(\omega)$ from
Eq. (\ref{DielConstJVL}),
\begin{equation}
  \frac{\varepsilon_{c}(\tilde{\omega})}{\varepsilon_{c}}\!\approx\!
   1\!+\!\frac{i\nu_{c}}{\tilde{\omega
}}\!-\!\frac{2}{h^{2}\tilde{\omega}^{2}}\!-\!\frac{1/(2\tilde{\omega}^{2})}{\tilde{\omega}^{2}\!+\!i\nu_{c}
\tilde{\omega}\!-\!\left(  1\!-\!i\nu_{ab}\tilde{\omega}\right)
h^{2}\!/4}.\label{DielCnstHighH}
\end{equation}
In the low-damping regime $\nu_{c},\nu_{ab}\ll1$ the loss function
has a peak at $\tilde{\omega}=h/2$ (in real units
$\omega=\omega_{p}\pi\gamma s^{2} B_{y}/\Phi_{0}$) corresponding to
the homogeneous plasma mode.\cite{BulPRB97} Such linear growth of
the plasma frequency with field has been indeed observed
experimentally in underdoped BSCCO by Kakeya \textit{et al.}
\cite{KakeyaPRB05}

\begin{figure}[ptb]
\begin{center}
\includegraphics[width=3.4in]{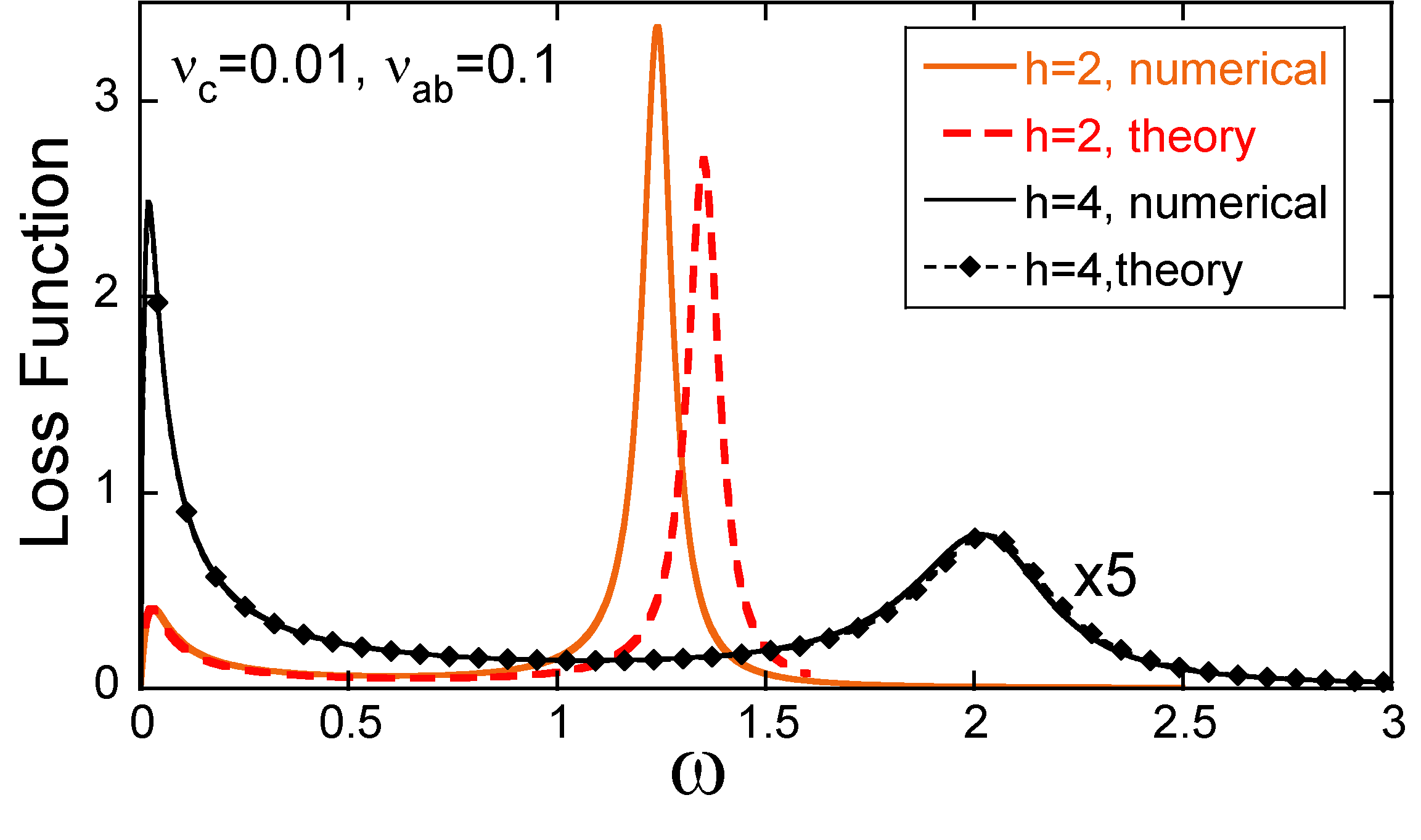}
\end{center}
\caption{(Color online) Comparison between the approximate
high-field limit of the loss function based on Eq.\
(\ref{DielCnstHighH}) and exact numerical solution for two field
values, $h=2$ and $4$ in the case of weak dissipation, $\nu
_{c}=0.01$ and $\nu_{ab}=0.1$. For clarity, the vertical scale for
the $h=4$ plots is magnified 5 times.} \label{Fig-LossHigh_hCompare}
\end{figure}

To verify the accuracy of the high-field approximation (\ref{DielCnstHighH}),
we compare in Fig.\ \ref{Fig-LossHigh_hCompare} the loss function obtained from
this formula with an accurate numerical solution for two field values, $h=2$
and $4$ in the case of weak dissipation, $\nu_{c}=0.01$ and $\nu_{ab}=0.1$. One
can see that the high-field formula accurately describes the low-frequency
region already at $h=2$, but it overestimates the peak frequency. At $h=4$ the
high-frequency approximation is already undistinguishable from the exact
result.

\section{Vortex-oscillations regime at small fields and frequencies
\label{sec:VortOscModel}}

In this section we consider the phenomenological theory of vortex oscillations
which describes the response of the vortex lattice at small frequencies,
$\omega\ll\omega_{p}$. For the Abrikosov vortex lattice such a theory was
developed by Coffey and Clem.\cite{CoffeyClemPRB92} The dynamic dielectric
constant for the Josephson vortex lattice has been derived using a similar
approach by Tachiki, Koyama, and Takahashi.\cite{TKTPRB94}

Consider a superconductor in the vortex state carrying ac supercurrent
$j_{s}\propto\exp(-i\omega t)$ along the $c$ axis. The ac electric field
consists of the London term and the contribution from the vortex oscillations,
\begin{equation}
E_{z}=-\frac{4\pi\lambda_{c}^{2}}{c^{2}}i\omega
j_{s}-\frac{B_{y}}{c}i\omega u.\label{ElVort}
\end{equation}
The vortex oscillation amplitude, $u$, can be found from the
equation
\begin{equation}
\left(  -\rho_{J}\omega^{2}-i\eta_{J}\omega+K\right)
u=\frac{\Phi_{0}} {c}j_{s},\label{VortOscEq}
\end{equation}
where $\rho_{J}$ is the linear mass of the Josephson vortex
\cite{CoffeyClemVMassPRB92}, $\eta_{J}$ \ is its viscosity coefficient
\cite{CoffeyClemJViscPRB90,JVflux-flow}, and $K$ is the spring constant due to
pinning (it is neglected in the rest of the paper). In Appendix
\ref{App:VortMassVisc} we present formulas for $\rho_{J}$ and $\eta_{J}$ in
terms of superconductor parameters. Finding the oscillating amplitude, we
obtain
\begin{equation}
E_{z}=-\left(  4\pi\lambda_{c}^{2}+\frac{B_{y}\Phi_{0}}{-\rho_{J}\omega
^{2}-i\eta_{J}\omega+K}\right)  \frac{i\omega j_{s}}{c^{2}}.\label{EzVortOsc}
\end{equation}
A finite electric field also generates the quasiparticle current
$j_{n} =\sigma_{c,n}(\omega)E_{z}$. Therefore, the total
conductivity $\sigma _{c}(B_{y},\omega)=(j_{s}+j_{n})/E_{z}$ is
given by
\begin{align}
\sigma_{c}(B_{y},\omega)= &  \sigma_{c,n}(\omega)\nonumber\\
- &  \frac{\varepsilon_{c}\omega_{p}^{2}}{4\pi i\omega}\left(
1\!+\!\frac
{1}{-\!\rho_{J}\omega^{2}\!-\!i\eta_{J}\omega\!+\!K}\frac{B_{y}\Phi_{0}}
{4\pi\lambda_{c}^{2}}\right)  ^{-1}.\label{CondVortOsc}
\end{align}
The interesting feature of the real part of conductivity,
\begin{align}
\sigma_{c,1}(B_{y},\omega)  & =\operatorname{Re}[\sigma_{c,n}(\omega
)]\nonumber\\
& +\frac{\eta_{J}\varepsilon_{c}\omega_{p}^{2}B_{y}\Phi_{0}/\left(
4\pi\lambda_{c}\right)  ^{2}}{\left(
K\!+\!B_{y}\Phi_{0}/(4\pi\lambda_{c}
^{2})-\!\rho_{J}\omega^{2}\right)  ^{2}\!+\!\left(
\eta_{J}\omega\right)
^{2}},\label{RealConVortOsc}
\end{align}
is a resonance peak at the frequency $\omega_{r}=\sqrt{\left[ K\!+\!B_{y}
\Phi_{0}/(4\pi\lambda_{c}^{2})\right]  /\rho_{J}}$. As one can see from Eq.
(\ref{EzVortOsc}), this resonance is a result of compensation of the London and
vortex contributions to the oscillating electric field at $\omega
\sim\omega_{r}$. Using the result for the vortex linear mass $\rho_{J}$ from
Appendix \ref{App:VortMassVisc} and neglecting pinning, the formula for
$\omega_{r}$ can be rewritten more transparently as $\omega_{r}\approx
0.84\omega_{p}\sqrt{2\pi\gamma s^{2}B_{y}/\Phi_{0}}$. At $\omega\rightarrow0$,
Eq. (\ref{RealConVortOsc}) gives a known result for the dc flux-flow
conductivity.\cite{JVflux-flow}

Using the well-known relation between the dynamic dielectric constant and
conductivity, $\varepsilon_{c}(\omega)=\varepsilon_{c}-4\pi\sigma_{c}
(\omega)/i\omega$, we obtain  \cite{TKTPRB94}
\begin{align}
\varepsilon_{c}(B_{y},\omega)= &  \varepsilon_{c}\!-\!\frac{4\pi\sigma
_{c,n}(\omega)}{i\omega}\nonumber\\
&  -\frac{\varepsilon_{c}\omega_{p}^{2}}{\omega^{2}}\left(  1\!+\!\frac
{1}{-\rho_{J}\omega^{2}\!-\!i\eta_{J}\omega\!+\!K}\frac{B_{y}\Phi_{0}}
{4\pi\lambda_{c}^{2}}\right)  ^{-1}.\label{DielConstVrt}
\end{align}
Using again formulas for $\rho_{J}$ and $\eta_{J}$ from Appendix
\ref{App:VortMassVisc} and neglecting pinning, we rewrite this result in
reduced variables in the form, convenient for comparison with numerical
calculations,
\begin{equation}
\frac{\varepsilon_{c}(h,\tilde{\omega})}{\varepsilon_{c}}=
 1\!+\!\frac{i\nu_{c}}{\tilde{\omega}}
\!-\!\frac{1/\tilde{\omega}^{2}}{1\!-\!2\pi h/[  C_{c}(
\tilde{\omega}^{2} \!+\!i\nu_{c}\tilde{\omega})
\!+\!C_{ab}i\nu_{ab}\tilde{\omega}] } \label{LossVrtRed}
\end{equation}
with $C_{c}\approx9.0$ and $C_{ab}\approx2.4$. We remind that this
formula is valid only at small fields, $h\ll1$, and frequencies,
$\tilde{\omega}\ll 1$.

\section{Field evolution of the Loss Function}

\begin{figure}[ptb]
\begin{center}
\includegraphics[width=2.8in]{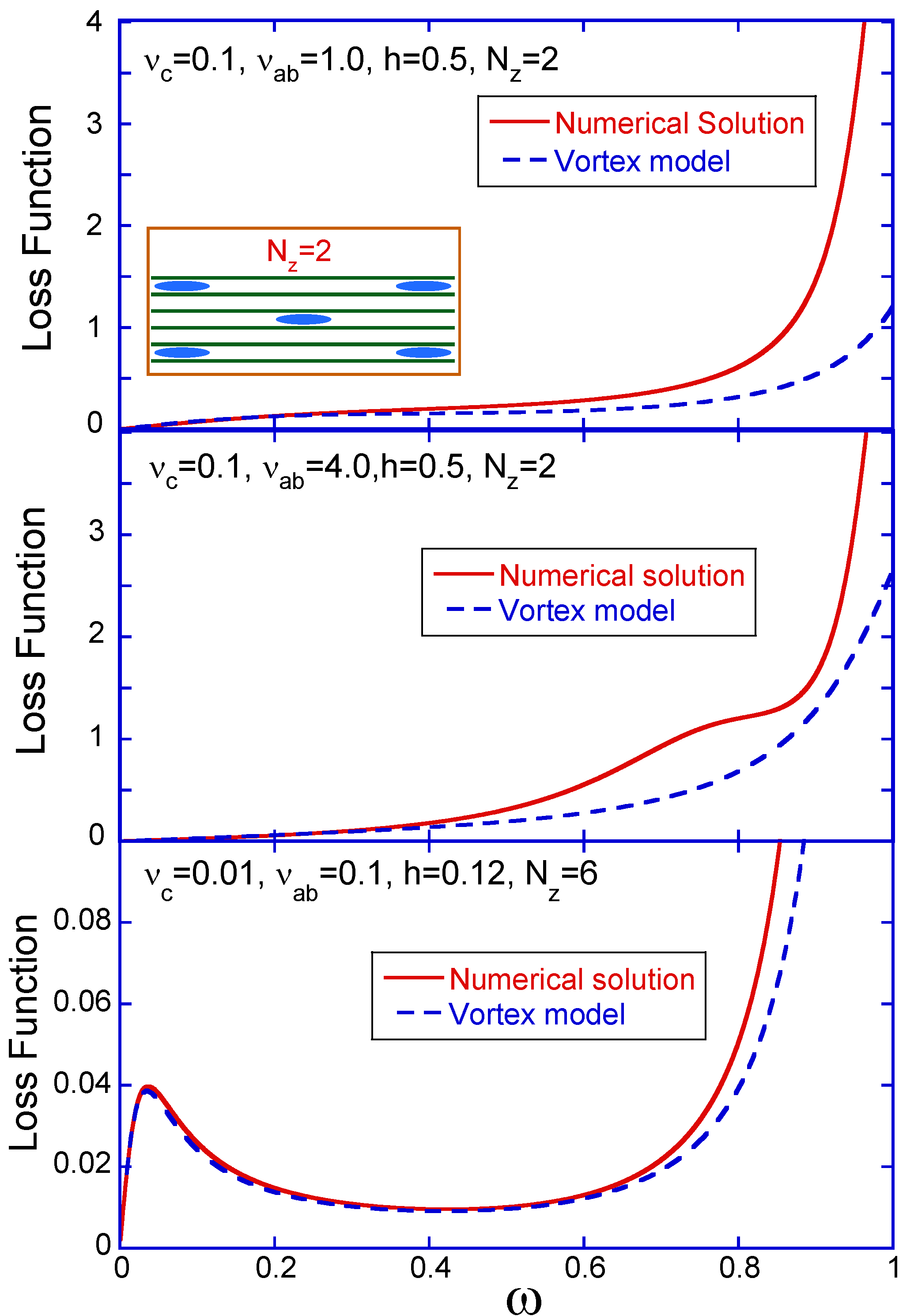}
\end{center}
\caption{(Color online) Comparison between the vortex-oscillation
model and exact numerical calculation for the loss function at
several representative values of the dissipation parameters, field,
and $N_{z}$ (shown in the plots). One can see that this model
typically describes the high-frequency response roughly up to half
of the plasma frequency. The inset in the upper plot illustrates the
aligned lattice with $N_{z}=2$.} \label{Fig-LossVortCompare}
\end{figure}

In the full range of frequencies and fields, the dynamic dielectric function
can only be computed numerically. At the first step, one has to find the static
phase distribution from Eq.\ (\ref{StatPhEq}) assuming a definite vortex
lattice structure. To probe general trends, we limit ourselves here only by
simple aligned lattices. At a fixed magnetic field, such a lattice is fully
defined by the number of layers separating the layers filled with vortices,
$N_{z}$, see sketch of the aligned lattice with $N_{z}=2$ in the inset of Fig.\
\ref{Fig-LossVortCompare}. At small fields, such lattices are realized in
ground states in the vicinity of two sets of commensurate fields, $B_{1}
(N_{z})=\sqrt{3}\Phi_{0}/(2N_{z}^{2}\gamma s^{2})$ and $B_{2} (N_{z})=\Phi
_{0}/(2\sqrt{3}N_{z}^{2}\gamma s^{2})$ ($h_{1}(N_{z})=\pi\sqrt{3}/N_{z}^{2}$
and $h_{2}(N_{z})=\pi/(\sqrt{3}N_{z}^{2})$ in reduced units). At the
intermediate field values the ground state is given by misaligned
lattices.\cite{KoshGrState,NonomHuPRB06}

At the first stage, we solved the static phase equations (\ref{StatPhEq}) for
fixed $h$ and $N_{z}$. This solution has been used as an input for the dynamic
phase equations (\ref{DynPhEq}). Finally, the oscillating phase determines the
dynamic dielectric constant (\ref{DielConstJVL}) and the reduced loss function
$L(\tilde{\omega})=-\mathrm{Im}[\varepsilon_{c}/\varepsilon_{c}(\tilde{\omega
})]$.

\begin{figure}[ptb]
\begin{center}
\includegraphics[width=3.4in]{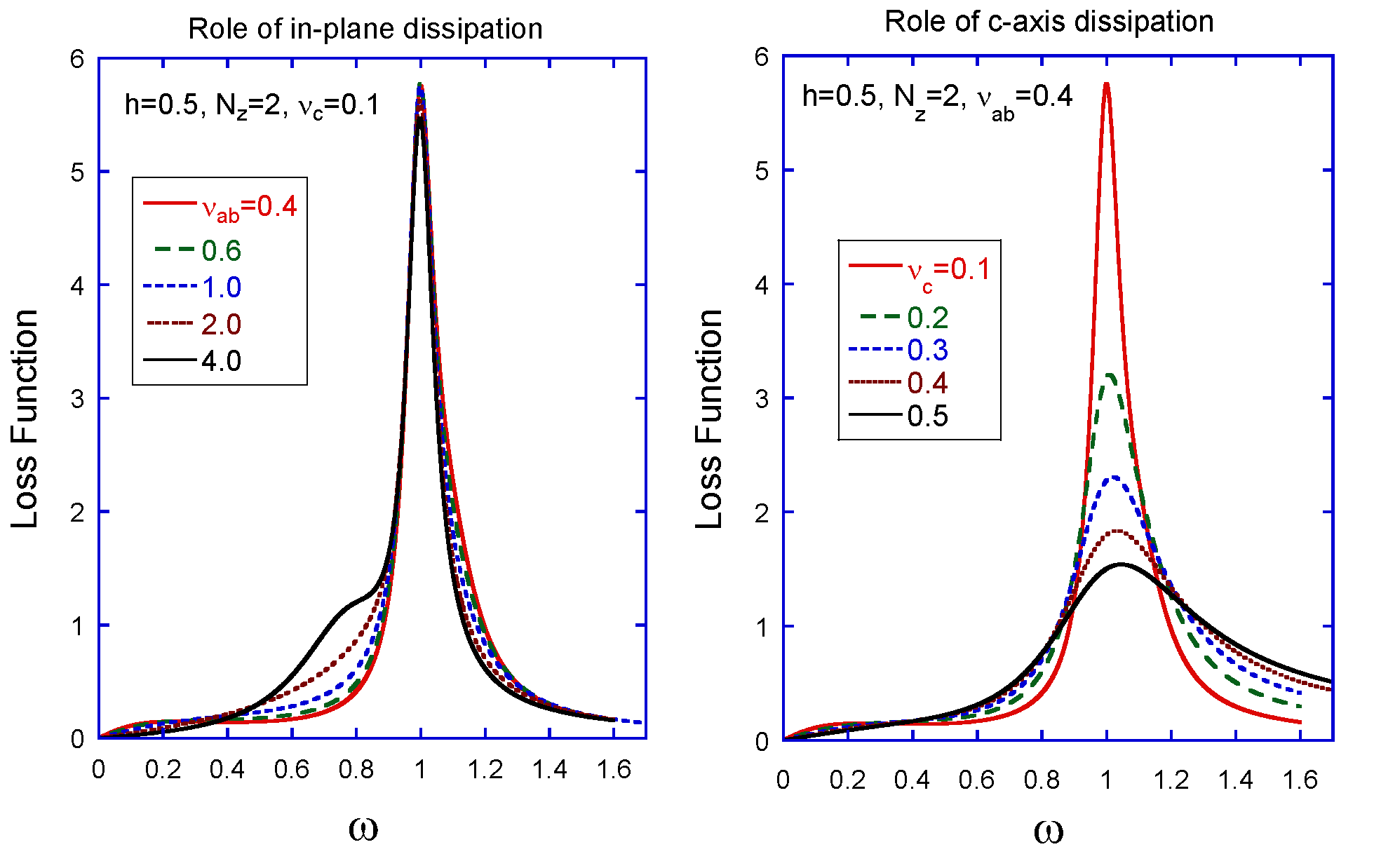}
\end{center}
\caption{(Color online) Influence of the dissipation parameters on
the loss-function shape in the dilute-lattice regime at $h=0.5$ and
$N_{z}=2$. In-plane dissipation (left plot) does not influence much
the JPR peak, but strongly influences the shape below the peak.}
\label{Fig-RoleDissip}
\end{figure}
We start with the discussion of several general properties of the
high-frequency response. To illustrate the roles of two different dissipation
channels, we show in Fig.\ \ref{Fig-RoleDissip} series of the frequency
dependencies of the loss function when only one damping constant is changed for
fixed field $h=0.5$ and $N_{z}=2$ corresponding to a dilute lattice. We can see
that the width of the JPR peak is mainly determined by the c-axis dissipation,
while the in-plane dissipation has no visible influence on the loss function
near the JPR peak. On the other hand, the in-plane dissipation strongly
influences the shape of the loss function below the peak. In particular, at a
very strong in-plane dissipation a peaklike feature appears below the JPR
frequency which looks like an oscillation mode.

In Fig.\ \ref{Fig-LossVortCompare} we compare the behavior of the numerically
computed loss function at low frequencies with predictions of the
vortex-oscillation model described in section \ref{sec:VortOscModel}. We can
see that this model typically accurately describes the behavior at frequencies
smaller then half the plasma frequency. At low dissipation, the loss function
has an additional peak at small frequencies. This peak disappears with
increasing dissipation. We have to emphasize that these plots are made for an
ideal homogeneous system. In a real layered superconductor, pinning of the
Josephson vortices by inhomogeneities will lead to the resonance peak at the
pinning frequency.

\subsection{Small dissipation (BSCCO)}

\begin{figure*}[ptb]
\begin{center}
\includegraphics[width=5in]{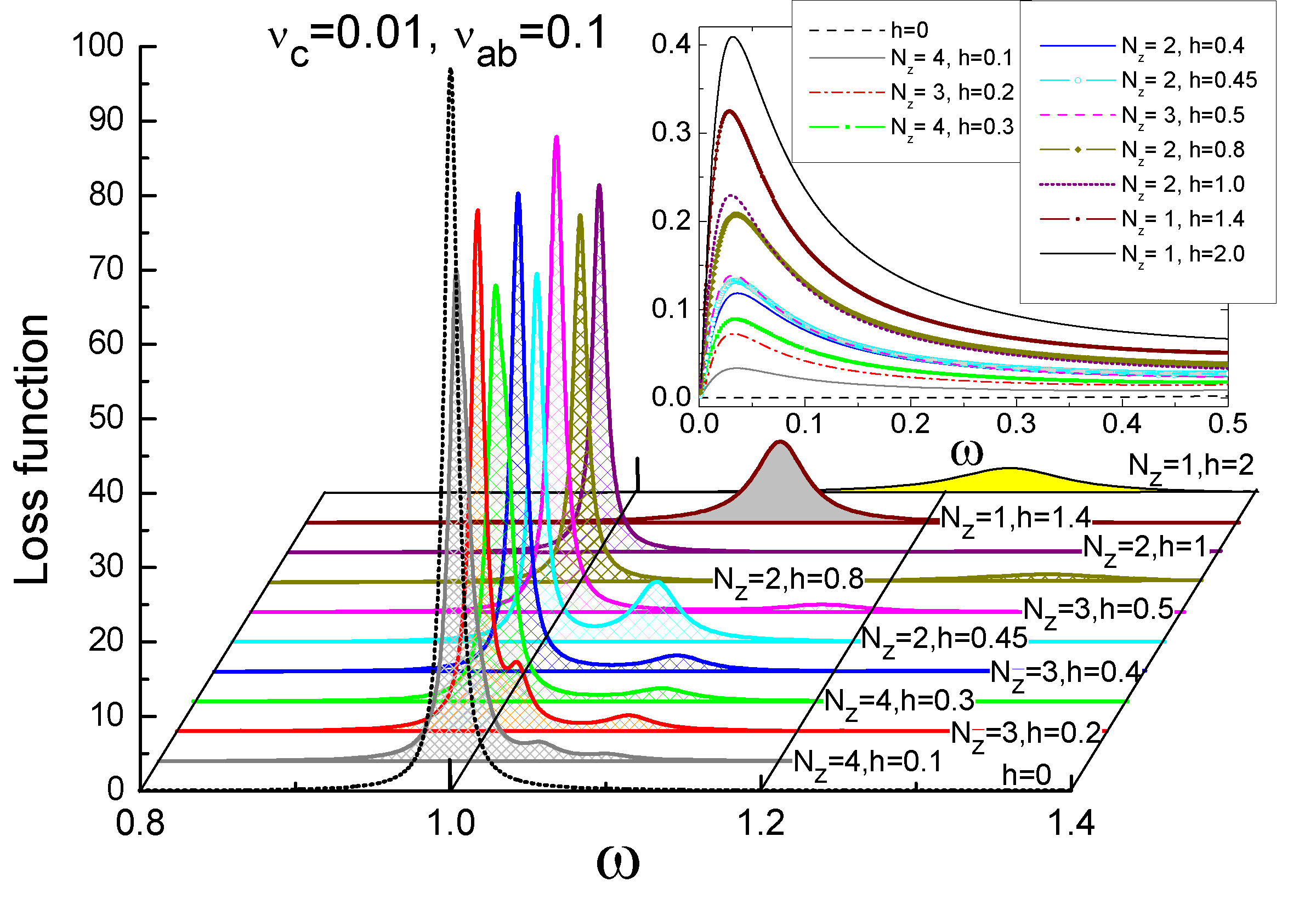}
\end{center}
\caption{(Color online) Series of the frequency-dependent loss
functions near JPR frequency at different fields for $\nu_{c}=0.01$
and $\nu_{ab}=0.1$. The inset shows the low-frequency region for the
same set of parameters.} \label{Fig-Loss_vc0_01vab0_1}
\end{figure*}

To illustrate a theoretically expected behavior of the high-frequency response
in BSCCO, we made calculations with small dissipation parameters, typical for
this compound, $\nu_{c}=0.01$ and $\nu_{ab}=0.1$. Figure
\ref{Fig-Loss_vc0_01vab0_1} shows the evolution of the loss function with
increasing magnetic field in the frequency range near the plasma resonance. For
such small dissipation, the loss function at zero field has a very sharp peak
at the plasma frequency. In the dilute-lattice regime at small fields, this
peak decreases in amplitude and is displaced to slightly lower frequencies.

\begin{figure}[ptb]
\begin{center}
\includegraphics[width=3.4in]{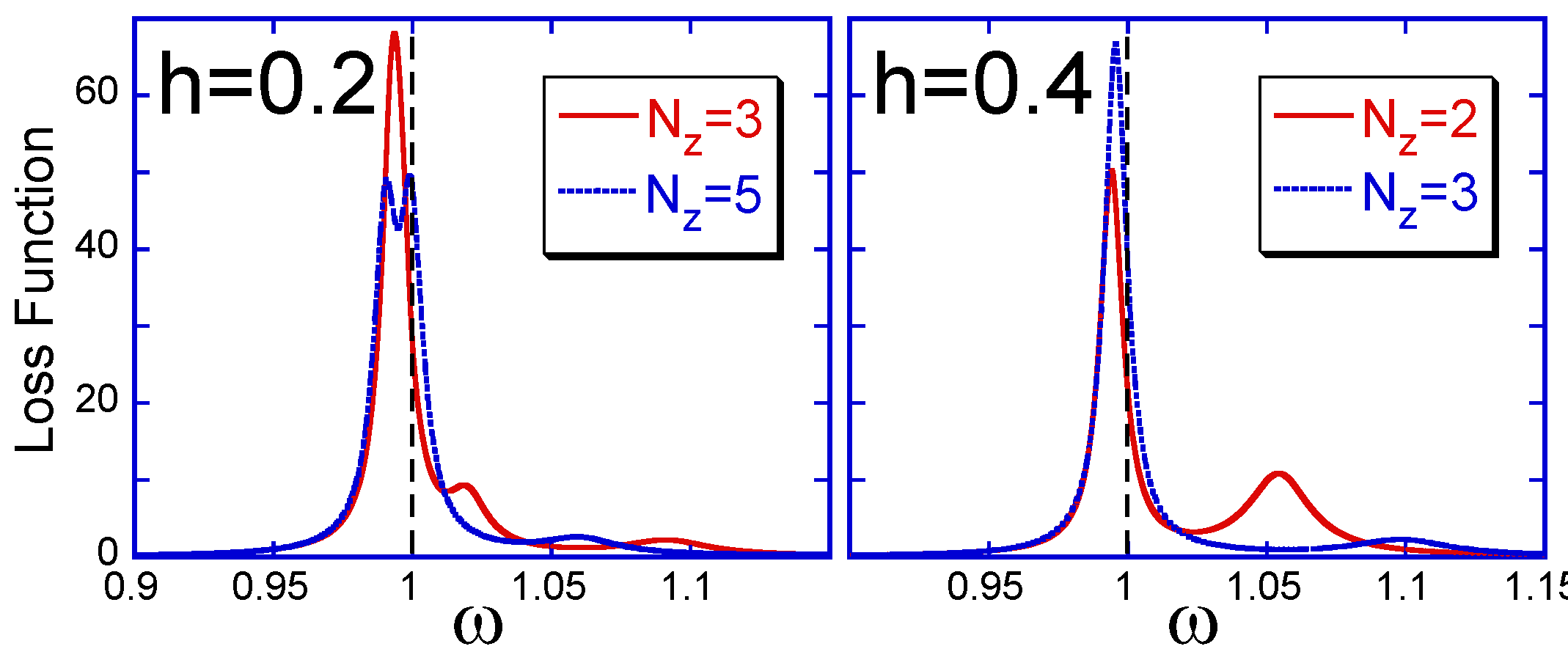}
\end{center}
\caption{(Color online) Plots of the loss function at the same $h$ but with
different $N_z$ for $\nu_{c}=0.01$ and $\nu_{ab}=0.1$. This figure illustrates
the sensitivity of the loss-function shape to the vortex lattice structure
(value of $N_{z}$) above the JPR peak.} \label{Fig-LossNzDep}
\end{figure}

The most interesting feature of the dilute regime is the appearance of the
satellite peaks in the high-frequency part. The strongest satellite is observed
at $N_{z}=2$ near $h\sim0.45$. A physical origin of these satellites is clear.
In the vortex lattice state, the homogeneous oscillations are coupled to the
plasma modes with the wave vectors set by the lattice. Therefore, the location
of the satellites is determined by the plasma spectrum.  For reference, we
present the plasma frequencies at the reciprocal-lattice vectors in Appendix
\ref{App:Satel}. At small dissipations the shape of the loss function is not
rigidly fixed by the value of the magnetic field, it is also sensitive to the
lattice structure. This is illustrated in Fig.\ \ref{Fig-LossNzDep}, where the
loss function is plotted at fixed $h$ for two values of $N_{z}$, for which the
lattice energies are close. One can see that the loss function is almost
$N_{z}$ independent at frequencies below the JPR peak indicating that in this
frequency range the vortices contribute independently to the response. However,
above the peak the shape of the loss function is very sensitive to $N_{z}$.
This means that the high-frequency response can potentially be used to probe
the lattice structure.

In the dense-lattice limit, the plasma peak moves to a higher frequency and its
intensity rapidly decreases, as predicted by the analytical theory in Section
\ref{sec:HighField}. This behavior is also consistent with experiment
\cite{KakeyaPRB05}. The transition to the dense lattice is very distinct. While
for $N_{z}=2$ and $h=1$ the peak is still located very close to the zero-field
JPR frequency and its amplitude is comparable with the zero-field peak, at very
close field $h=1.4$ in the dense-lattice regime, $N_{z}=1$, the peak is
noticeably shifted to a higher frequency, its amplitude considerably dropped,
and the width increased.

An additional broad peak exists at low frequencies. Its intensity is much
smaller than the JPR peak. The evolution of this peak with increasing magnetic
field is shown in the inset of Fig.\ \ref{Fig-Loss_vc0_01vab0_1}. One can see
that the intensity of this peak monotonically increases with the magnetic
field. This peak is well described by the vortex-oscillation model.

We did not find any intrinsic resonances in the loss functions  at frequencies
smaller than the JPR frequency, meaning that there are no modes coupled to
homogeneous oscillations in this frequency range for a homogeneous
superconductor.\cite{AntiphaseMode} This suggests that the resonance feature
observed in underdoped BSCCO at $\sim0.5 \omega_{p}$ in Ref.\
\onlinecite{KakeyaPRB05} probably has an extrinsic origin, e.g., it may be
caused by the pinning of the Josephson vortices.

\subsection{Large dissipation (underdoped YBCO)}

\begin{figure}[ptb]
\begin{center}
\includegraphics[width=3.4in]{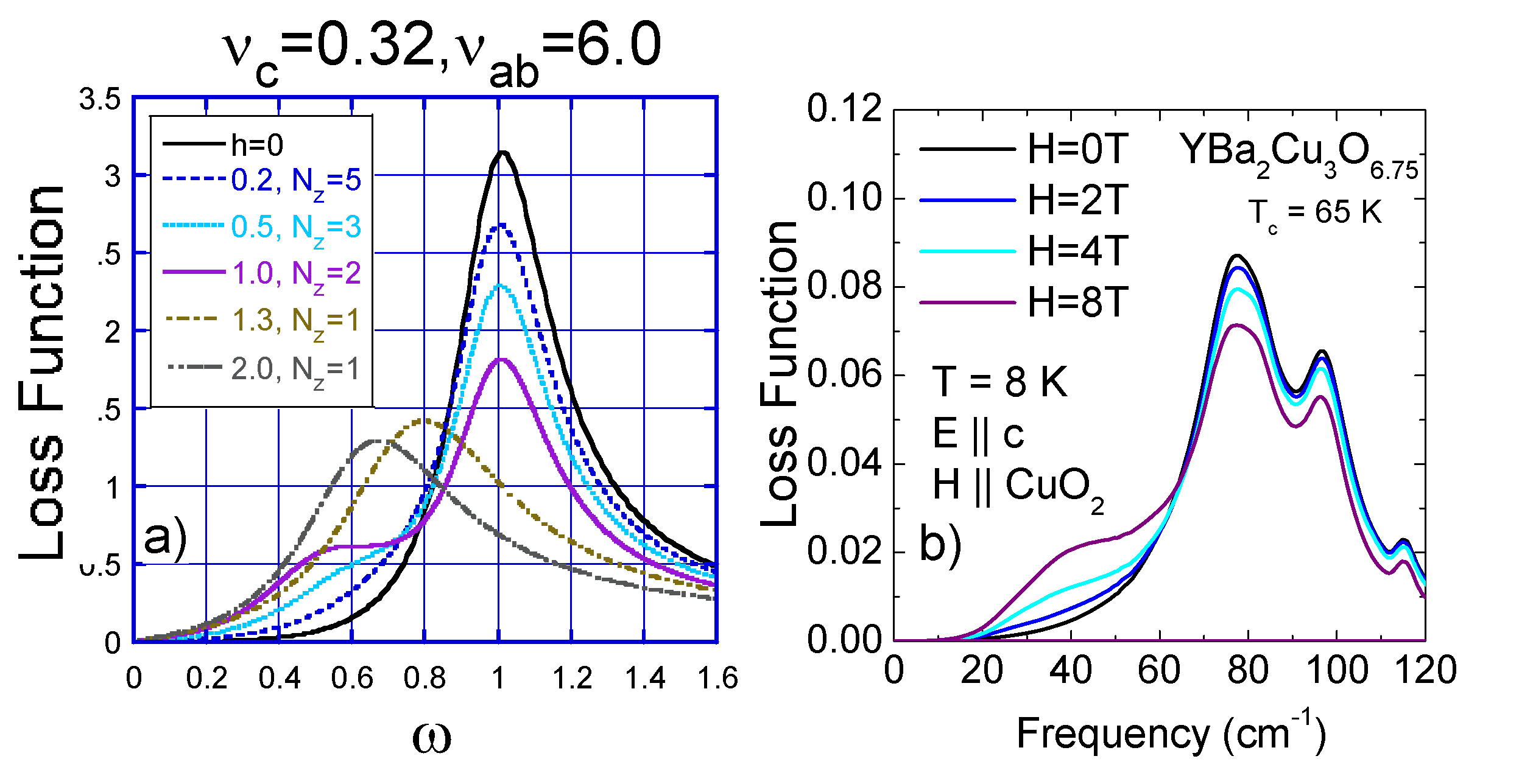}
\end{center}
\caption{(Color online) (a)Series of the frequency-dependent loss functions at
different fields for $\nu_{c}=0.32$ and $\nu_{ab}=6.0$. For comparison plot (b)
shows  the evolution of the loss function with increasing in-plane field for
underdoped YBCO from Ref.\ \onlinecite{LaForge}.} \label{Fig-Loss_vc0_32vab6_0}
\end{figure}\begin{figure}[ptb]
\begin{center}
\includegraphics[width=2.8in]{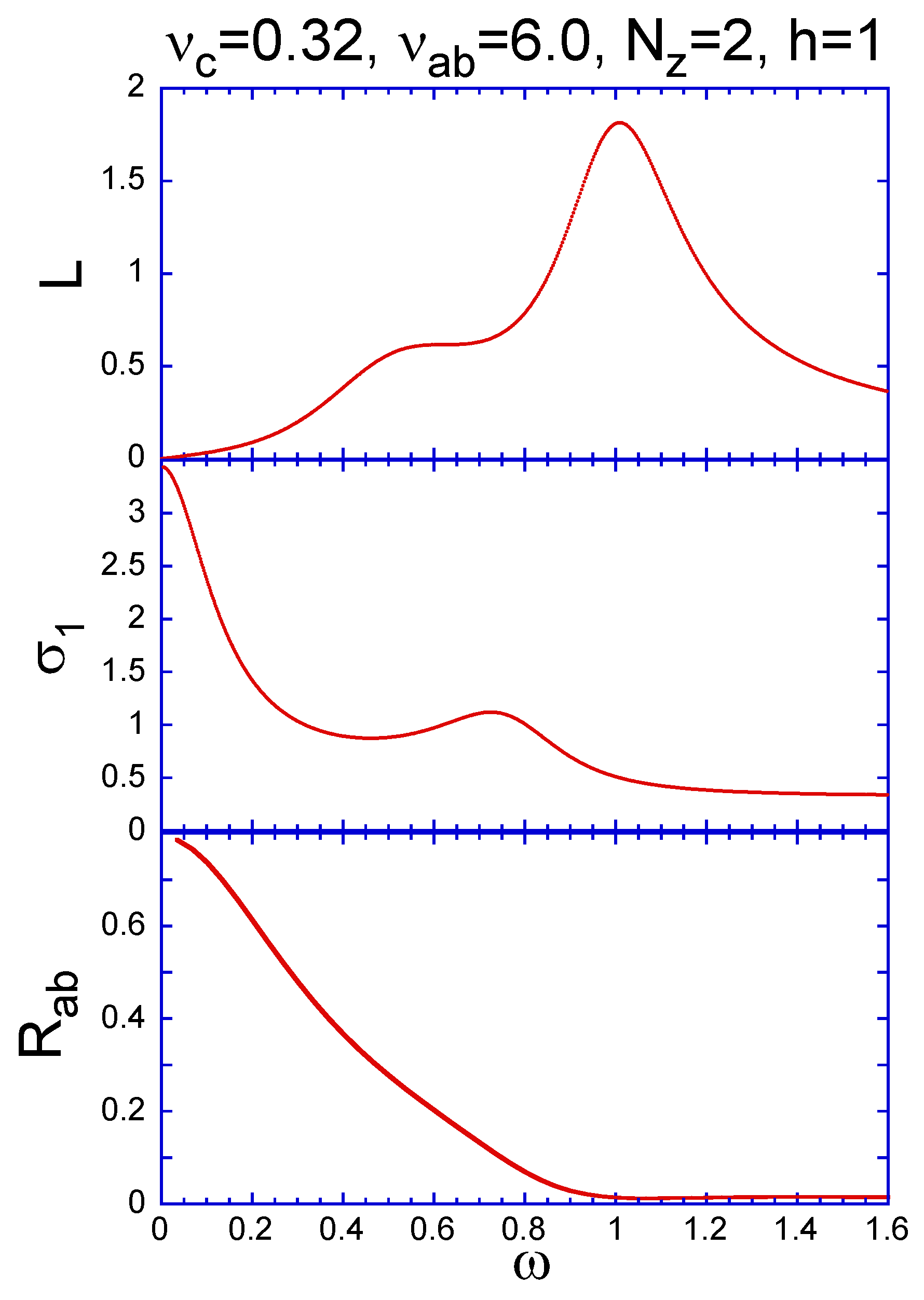}
\end{center}
\caption{(Color online) The frequency dependences of the loss function, real
part of conductivity, and relative contribution of in-plane dissipation
(\ref{Rab}) for the same damping parameters as in Fig.\
\ref{Fig-Loss_vc0_32vab6_0}, $N_{z}=2$, and $h=1$. Rapid decreasing of in-plane
inhomogeneity of the oscillations with increasing frequency can also be seen
from the visualization of oscillating patterns of the local electric fields at
different frequencies.\cite{Anim}} \label{Fig-LossRDissvc0_32vab6_0}
\end{figure}
In this subsection we discuss the high-frequency response of the Josephson
vortex lattice in the case of large effective dissipation, which is realized,
e.g., for the underdoped YBCO.\cite{KojimaJLTP03,LaForge} Figure
\ref{Fig-Loss_vc0_32vab6_0} shows the frequency dependences of the loss
functions for the dissipation parameters $\nu_{c}=0.32$ and $\nu_{ab}=6.0$. For
comparison, we also present the loss functions at different in-plane fields for
an underdoped YBCO sample from Ref.\ \onlinecite{LaForge}. One can see that two
experimental features are well reproduced by the theory: decreasing intensity
of the main peak with increasing field and the appearance of the satellite peak
in the low-frequency part of the line. The origin of this satellite peak is
distinctly different from the high-frequency satellites found for small
dissipation. Figure\ \ref{Fig-RoleDissip} demonstrates that this peak only
appears for sufficiently strong in-plane dissipations. As this peak is absent
in the low-dissipation limit, it does not correspond to any specific mode of
phase oscillations.  It is caused by the decrease of the relative contribution
of the in-plane dissipation channel with increasing frequency. To verify this
interpretation, we plot in Fig.\ \ref{Fig-LossRDissvc0_32vab6_0} the loss
function for $N_{z}=2$ and $h=1$ together with the relative contribution of
in-plane dissipation to losses, $R_{ab}= \sigma_{ab} \overline{E_{x}^{2}} /(
\sigma_{c}\overline{E_{z}^{2}}+\sigma_{ab}\overline{E_{x}^{2}} )$, which can be
rewritten in reduced coordinates,
\begin{equation}
R_{ab}=\frac{\nu_{ab}\overline{\left\vert \partial\phi_{n}/\partial
x\right\vert ^{2}}}{\nu_{c}\left(  1+\overline{|\phi_{n+1}
-\phi_{n}|^{2} }\right) +\nu_{ab}\overline{\left\vert
\partial\phi_{n}/\partial x\right\vert
^{2}} }.\label{Rab}
\end{equation}
One can see that $R_{ab}$ decreases with increasing frequency and almost
vanishes above the plasma resonance. The small-frequency peak roughly
corresponds to the frequency where $R_{ab}$ drops to one half. The decrease of
the in-plane dissipation also leads to the peak in the real part of the optical
conductivity $\sigma_1$, also shown in Fig.\ \ref{Fig-LossRDissvc0_32vab6_0}.
Note that the peak in $\sigma_1$ corresponds to the dip in the loss function.

At high fields, in the dense-lattice limit, only a single broad dissipation
peak remains. In contrast to the low-dissipation case, this peak is displaced
to lower frequencies with increasing magnetic field. This behavior is not
verified experimentally yet, data are only available for field range
$h\lesssim1$.\cite{KojimaJLTP03,LaForge}

\section{Summary}

In summary, we developed a comprehensive description of the high-frequency
response of the Josephson vortex lattice in layered superconductors. We found
the general relation (\ref{DielConstJVL}) connecting the dynamic dielectric
constant with the averages containing the static and oscillating phases.
Analytical formulas for the dynamic dielectric constant and loss function were
derived for the high-field regime and the vortex-oscillation regime at low
frequencies. Numerically solving equations for the oscillating phases, we
explored the evolution of the loss function with increasing magnetic field for
the cases of weak dissipation describing BSCCO and strong dissipation
describing underdoped YBCO. Several new features were found. The most
interesting feature in the weak-dissipation case is the high-frequency
satellites in the dilute-lattice regime, which appear due to the excitation of
plasma modes at the wave vectors of the reciprocal lattice. In the
strong-dissipation limit, we reproduced the additional peak in the loss
function below the JPR peak experimentally observed in underdoped
YBCO.\cite{LaForge,KojimaJLTP03} We established that this peak appears due to
the frequency dependence of the in-plane contribution to losses.

\section{Acknowledgments}

This work was motivated by illuminating discussions of experimental data with
D. Basov, A. LaForge, and I.\ Kakeya. This work was supported by the
U.\ S.\ DOE, Office of Science, under contract \# DE-AC02-06CH11357.

\appendix

\section{Josephson-vortex mass and viscosity \label{App:VortMassVisc}}

The viscosity of an isolated Josephson vortex has been considered in Refs.\
\onlinecite{CoffeyClemJViscPRB90} and \onlinecite{JVflux-flow}. In the case
when the dissipation is caused both c-axis and in-plane quasiparticle
transport, the viscosity coefficient is given by
\begin{align}
\eta_{J}  &  =\frac{1}{\gamma s^{2}}\left(  \frac{\Phi_{0}}{2\pi c}\right)
^{2}\left[  C_{c}\sigma_{c}+C_{ab}\frac{\sigma_{ab}}{\gamma^{2}}\right]
,\nonumber\\
&  =\frac{\varepsilon_{c}\omega_{p}\Phi_{0}^{2}}{\pi\left(  4\pi cs\right)
^{2}\gamma}\left[  C_{c}\nu_{c}+C_{ab}\nu_{ab}\right]  ,\label{JVVisc}
\end{align}
where the numerical constants $C_{c}$ and $C_{ab}$ are determined by the phase
distribution of an isolated Josephson vortex, $\phi_{n}^{(0)}$,
\begin{align*}
C_{c}  &  =\sum_{n=-\infty}^{\infty}\int_{-\infty}^{\infty}du\left(
\frac{\partial\left(  \phi_{n+1}^{(0)}-\phi_{n}^{(0)}\right)  }{\partial
u}\right)  ^{2}\approx9.0,\\
C_{ab}  &  =\sum_{n=-\infty}^{\infty}\int_{-\infty}^{\infty}du\left(
\frac{\partial^{2}\phi_{n}^{(0)}}{\partial u^{2}}\right)  ^{2}\approx2.4.
\end{align*}

The mass of the Josephson vortex has been considered in Ref.\
\onlinecite{CoffeyClemVMassPRB92}. This mass is determined by the kinetic
energy, $\mathcal{E}_{k}$, which for a moving Josephson vortex can be presented
as
\begin{align*}
\mathcal{E}_{k}  &  =\int d^{3}\mathbf{r}\frac{\varepsilon_{c}E_{z}^{2}}{8\pi
}=s\sum_{n}\int d^{2}\mathbf{r}\frac{\varepsilon_{c}}{8\pi}\left(  \frac
{\Phi_{0}}{2\pi cs}\right)  ^{2}\dot{\theta}_{n}^{2}\\
&  =\frac{v^{2}}{2}s\sum_{n}\int d^{2}\mathbf{r}\frac{\varepsilon_{c}}{4\pi
}\left(  \frac{\Phi_{0}}{2\pi cs}\right)  ^{2}\left(  \frac{d\theta_{n}^{(0)}
}{dx}\right)  ^{2}.
\end{align*}
For a slowly moving vortex with velocity $v$, the phase difference is
determined by its static distribution,
$\theta_{n}(\mathbf{r},t)=\theta_{n}^{(0)}(x-vt)$, and we obtain
\[
\mathcal{E}_{k}=L_{y}\frac{v^{2}}{2}s\sum_{n}\int dx\frac{\varepsilon_{c}
}{4\pi}\left(  \frac{\Phi_{0}}{2\pi cs}\right)  ^{2}\left(  \frac{d\theta
_{n}^{(0)}}{dx}\right)  ^{2}.
\]
As, by definition, $\mathcal{E}_{k}=L_{y}\rho_{J}v^{2}/2$, we obtain the
following result for the linear vortex mass
\begin{align}
\rho_{J}  &  =\frac{\varepsilon\Phi_{0}^{2}}{\pi\left(  4\pi c\right)  ^{2}
s}\sum_{n=-\infty}^{\infty}\int_{-\infty}^{\infty}dx\left(  \frac{d\theta_{n}
}{dx}\right)  ^{2}\nonumber\\
&  =\frac{C_{c}\varepsilon\Phi_{0}^{2}}{\pi\gamma\left(  4\pi cs\right)  ^{2}
}.\label{JVMass}
\end{align}
The reduced combination in the dynamic dielectric constant
(\ref{DielConstVrt}) can be represented as
\[
\frac{1}{\rho_{J}\omega^{2}+i\eta_{J}\omega}\frac{B\Phi_{0}}{4\pi\lambda
_{c}^{2}}=\frac{2\pi h}{C_{c}\left(  \tilde{\omega}^{2}+i\nu_{c}\tilde{\omega
}\right)  +C_{ab}i\nu_{ab}\tilde{\omega}}.
\]

\section{Plasma frequencies at reciprocal lattice vectors: Expected location of satellites
\label{App:Satel}}

The satellite peaks are expected approximately at the plasma frequencies for
the wave vectors of the reciprocal lattice. In the reduced coordinates, the
plasma spectrum at the zero magnetic field and without the charging effects is
given by
\begin{equation}
\omega_{p}^{2}(\mathbf{k},q)=1+\frac{k^{2}}{1/l^{2}+2(1-\cos q)}
\label{PlasmSpectr}
\end{equation}
where $\mathbf{k}$ and $q$ are the wave-vector components along and
perpendicular to the layers respectively. The reciprocal vectors of the vortex
lattice are given by
\[
(k_{(n,m)},q_{(n,m)})=(2\pi n/a,2\pi(m-n/2)/N_{z}),
\]
where $n$ and $m$ are the integer indices and $a$ is the vortex
separation, $a=2\pi/hN_{z}$. Neglecting the small factor $1/l^{2}$,
we obtain for the satellite frequencies
\begin{equation}
\omega_{s}^{2}(n,m)  \approx   1+\frac{\left(  hN_{z}n\right)
^{2}/2}{1-\cos\left[  2\pi\left( m-n/2\right)  /N_{z}\right]
}.\label{SatFreq}
\end{equation}
The most intense satellite is expected for the basic reciprocal
vector with indices $(n,m)=(1,0)$
\[
\omega_{s}^{2}(1,0)=1+\frac{\left(  hN_{z}\right)
^{2}/2}{1-\cos\left[ \pi/N_{z}\right]  }
\]
In particular, for $N_{z}=2$, this gives $\omega_{s}^{2}(1,0)=1+2h^{2}$.

\end{document}